\documentclass{mem}
\usepackage{natbib}\usepackage{txfonts}\usepackage{balance}
\usepackage{graphicx}
\usepackage[a4paper,breaklinks,dvipdfm]{hyperref}
\idline{75}{282}
\begin{document}
\def\teff{$T\rm_{eff }$}
\def\kms{$\mathrm {km s}^{-1}$}

\title{Reading the book: from ``chemical anomalies" to 
``standard composition" of globular clusters
}

   \subtitle{}

\author{
Angela \,Bragaglia\inst{1}
\and Eugenio Carretta\inst{1}
\and Raffaele Gratton\inst{2}
\and Valentina D'Orazi\inst{3,4}
\and Sara Lucatello\inst{2}
\and Chris Sneden\inst{5}
          }

  \offprints{A. Bragaglia}

\institute{
Istituto Nazionale di Astrofisica --
Osservatorio Astronomico di Bologna, via Ranzani 1
I-40127 Bologna, Italy \email{angela.bragaglia@oabo.inaf.it
\and
Istituto Nazionale di Astrofisica --
Osservatorio Astronomico di Padova, vicolo Osservatorio 5
I-35122 Padova, Italy
\and
Department of Physics and Astronomy, Macquarie University, Balaclava Rd, North Ryde, NSW 2109, Australia
\and
Monash Centre for Astrophysics, School of Mathematical Sciences, Building 28, Monash University, VIC 3800, Australia
\and
Department of Astronomy and McDonald Observatory, The University of Texas, Austin, TX 78712, USA
}
}

\authorrunning{Bragaglia }

\titlerunning{Chemical composition of GCs}

\abstract{
It is now commonly accepted that globular clusters (GCs) have undergone a
complex formation and that they host at least two stellar generations. 
This is a recent paradigm and is founded on both photometric and
spectroscopic evidence.  We concentrate on results based on high-resolution spectroscopy and on
how we moved from single to multiple stellar populations concept for GCs. We underline that the peculiar chemical composition of GC
stars is fundamental in establishing the multiple populations scenario and 
briefly outline what can be learned from observations. Finally, recent
observational results on large samples of stars in different evolutionary phases
are discussed.

}
\maketitle{}

\begin{table*}
\caption{Which elements show variations and (anti-)correlations and where (see
\citealt{gratton12rev} for updated references). }
\label{tab1}
\begin{center}
\begin{tabular}{lll}
\hline
Elements & Where & Notes\\
\hline
C, N   & field stars and all GCs & C,N variations in GCs have primordial \\
       &                         & ~variation in addition to evolutionary one \\
O, Na  & (almost) all GCs        & see Fig.~1 for exceptions\\
Mg, Al & some GCs                & metal-poor and/or massive GCs \\
Si     & some GCs                & metal-poor and/or massive GCs \\
Li, F      & all (?) GCs             & very few GCs studied \\
He     & all (?) GCs             & small variation in most GCs, large variation \\
       &                         & ~in a few (e.g., $\omega$ Cen, NGC~2808)\\
n-capture & in a few GCs         & usually if also Fe varies?\\
Fe     & in a few GCs            & $\omega$ Cen, Ter 5, M22, M54, NGC~1851 \\
K      & NGC~2419                & varies much less in other GCs, from a few data\\
\hline
\end{tabular}
\end{center}
\end{table*}

\section{Chemistry and the nature of globular clusters}
We presently understand that globular clusters (GCs) are more complex than believed in the past. GCs do not host a single coeval, chemically homogeneous population
but show clear evidence of at least two generations, slightly different in age but having (very) different chemical composition. This is manifest from the colour-magnitude diagrams if the photometry is very precise and/or obtained with filters able to enhance the differences (e.g., \citealt{dantona05,piotto07} for NGC~2808, \citealt{milone1851,han} for NGC~1851) and more generally from the chemistry \cite[e.g.,][]{gratton12rev}. 
The topic is huge, so we will limit the discussion to the most recent results
in the field of our main expertise, i.e., high-resolution spectroscopy.
Thus, even if the first variations were noticed in the very light elements C and N, we will leave them aside (as we will do for Li and F) and  concentrate here on other elements. 

There are several chemical
elements that present star-to-star variations in
GCs (see Table~\ref{tab1}). These variations are not random, rather
some elements are found depleted in some stars when other are found enhanced
in the same stars. 
Summarizing, in a typical GC we have {\bf a)} stars with ``normal" He, C, O, Mg, Na, and Al (first-generation stars) and {\bf b)} stars with depleted C, O and
enhanced He, N, Na (second-generation stars). Sometimes, {\bf b1)} also depleted Mg and enhanced Al are seen, and/or {\bf b2)} He may be strongly enhanced.
{\bf c)} Very rarely, Fe or heavier elements present significant variations.

\subsection{The classical couples: Na and O, Al and Mg}
Maybe the most famous relation is between Na and O. These elements 
vary in step, with O being depleted and Na enhanced, and this happens only in GC stars \citep[see][where the work of the Lick-Texas group, who pioneered the field, is described]{araa}. The
finding of an Na-O anti-correlation also in non-evolved stars (\citealt{gratton01,rc02} and more authors later) has been the final
piece of evidence that these variations are not evolutionary, but have been
imprinted in the stars at birth and are due to the contribution of a previous
stellar generation. In fact low-mass, non-evolved stars are not able to
sinthesise Na (their core does not reach the very high
temperatures required to produce it through H-burning) and bring it to the surface. 

Recent work on the Na-O anti-correlation has benefited from the high-multiplexing, high-sensitivity, high-resolution spectrographs on large telescopes. It has been possible to study the phenomenon in large samples of stars -of the order of 100 stars per cluster- in many GCs with very precise measurements, so that quantitative analyses and comparisons could be made.
This has been done for stars on the red giant  \citep[RGB, see e.g.,][for M4, 15 and 17 GCs, and M13, respectively]{marino08,carretta09a,carretta09b,jp12}, main sequence  (MS, see e.g., \citealt{carretta04,dorazi10}), and horizontal branch  (HB, see e.g., \citealt{gratton11,gratton13,marino11c}). It is very interesting to see that the Na-O anti-correlation, while always present, is different from cluster-to-cluster, having different extension and shape; these differences seem correlated with general clusters properties (see \citealt{zcorri}).

At temperatures even higher than those at which Na is produced, Mg can be burned
into Al. However, this appears to happen only in a subsample of clusters; 
\cite{carretta09b} studied 18 GCs, finding a significant Al enhancement in about
one half of them. Studying the way this happens  especially in the more
metal-poor or more massive GCs, can give us insights on the nature of the stars
that polluted the gas from which the second generation formed. We will come back
to Al below.

\subsection{As light as He}
What produces the light elements variations is H-burning at high temperature,
so it is simply natural that also He should vary. The influence of He abundance on
the evolutionary sequences, especially the HB and MS, can be seen using photometry \citep[e.g.,][]{bedin,dantona05,piotto07,sbordone}.
However, in some ``fortunate" cases, He abundance can be directly measured, not
only inferred from the comparison to models. Generally, He lines are visible in hot
stars, but \cite{dupree} and especially \cite{pasquini} were able to measure He
in RGB stars (in $\omega$ Cen and NGC~2808, respectively).  An easier
task would seem to measure He in HB stars, and this has been done in a few
cases (NGC~6752, M4,  NGC~1851, M5: \citealt{villanova09,villanova12,gratton11,gratton13}) but it requires the right temperature, high resolution spectra, and very high S/N, so it is a difficult task. However, observations confirm that stars assigned to the second generation on the basis of their Na and O abundances have higher He content than first-generation stars.

\subsection{Heavier elements}
Apart from $\omega$ Cen and Terzan~5 \citep{ferraro}, there are a few more GCs 
for which also iron and heavier elements vary from star-to-star, probable indication of a formation mechanism even more complex than for normal GCs. One classical example is M22, for which a demonstration of true Fe dispersion has only recently been produced (see e.g., \citealt{marino09,marino11b}, who showed the presence of two groups, with neutron-capture elements correlated with Fe abundance). Another case is M54, for which we \citep{carretta10b,carretta10c} showed that the Fe-poor and Fe-rich components have both an Na-O
anti-correlation and that they are different, similar to what is found in $\omega$ Cen \citep{jp10,marino11a}; the two clusters could have formed in a similar way and be in a different evolutionary stage. Finally, NGC~1851 has been the target of many recent studies, since it shows many interesting features, like a split subgiant branch \citep{milone1851}, possibly tied to variations in CNO content and/or to age difference  \citep[e.g.,][]{cassisi08,ventura09}, dispersion in Fe and the possibility that it originated from a merger \citep{carretta10d,carretta11}, the presence of enhanced n-capture elements correlated with the light elements \citep[e.g.,][]{yg08,yong09,carretta11}, etc. It is clearly a cluster that merits further attention.

\begin{figure}[t!]
\resizebox{\hsize}{!}{\includegraphics[]{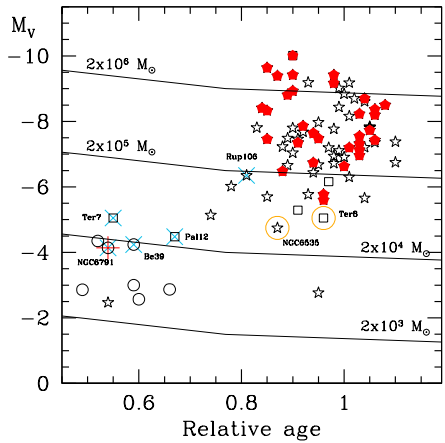}}
\caption{\footnotesize
Relative age parameter vs. absolute magnitude for globular
and old open clusters  (for a detailed description, see \citealt{be39}, from which the figure is taken). 
Red filled symbols are GCs where the Na-O
anti-correlation has been observed; open symbols stand for GCs not observed or  old open clusters. Light blue crosses indicate clusters that
do not show evidence of Na-O anti-correlation: two GC members of
Sagittarius dSph, Rup 106, and the two massive, old OCs, Be 39 and
NGC 6791 (the latter indicated also in red, since its situation is still
not completely assessed). Two orange, large circles indicate 
GCs with work on-going.
}
\label{figmvage}
\end{figure}

\section{The ``universality" of the Na-O anti-correlation in GCs}
We have seen that the Na-O anti-correlation seems a general
feature of GCs, so that we have proposed \citep{zcorri} that it could be used as a
sort of definition, indicating those clusters that are massive enough and were
born in an environment that favoured gas accumulation and production of a second
generation. This universality is shown in Fig.~\ref{figmvage}, where the separation between old, massive GCs, hosting multiple populations and younger, less massive open clusters (OCs), home of a single population, seems natural.
However, the region in-between still needs to be studied and we  have begun to investigate which is the lower mass at which multiple populations can manifest (recalling, however, the distinction between present-day and original cluster mass). Results are still not univocal for the old OC NGC6791, where \cite{geisler} find an Na-O anti-correlation while our work, still on-going, does not. They are instead very clear for Be~39, which we find very homogeneous and composed by a single population \citep{be39}.
Further clusters are under exam by our group or by others (e.g., Ter~8, in which
the Na-O anti-correlation, if
present, seems to be less extended than in more massive clusters, or
Rup~106, which does not show any) but the complete analyses are not published yet.

\section{Recent results and old questions}
An important diagnostic to understand what went on
at the GC formation and during the early evolution is the continuous or
discrete behaviour of properties. Photometrically, it has been demonstrated that
there are discrete distribution of He in MS or on the HB \citep[e.g.,][for NGC~2808 and $\omega$ Cen]{dantona05,piotto07,omega}, and discrete components
show up along other evolutionary sequences, especially when the right filter combinations are used \citep[e.g.,][]{milone12}. 

Spectroscopically, we need to obtain both large samples and precise measures
to check if distinct groups can be isolated by their chemical properties.
This has been possible for instance in M4, where \cite{marino08}
separated two distinct
groups in the Na and O distribution, and in NGC~2808, where we see three
groups, reminiscent of the tripartition of the HB and of the MS \citep{carretta06}.
A good element to study  seems to be Al. For instance, in NGC~6752 we were able to see three groups
in the ratio [Al/Mg]; they can be traced on the RGB using the Str\"omgren
index $c_y$, defined by \cite{yg} to measure N. So we can tie the CNO,
NeNa, and MgAl cycles together. For this cluster, the intermediate
population cannot be obtained diluting the two extremes, so it seems to require
two episodes of formation and two different first-generation polluters. 

Finally, we only briefly mention the recent finding of an anti-correlation between Mg and K in NGC~2419 \citep{cohen,mucciarelli}, a very enigmatic cluster. The extreme values found both for Mg and K seem to be an unicum, at least based on the scanty data available at the moment (Carretta et al., in prep.). Maybe this is an extreme manifestation of the usual Mg depletion \citep{paolo}, but it is certainly worth further investigation, to understand
what makes this cluster so peculiar.

We are learning every day more on the clusters' formation mechanism(s), but still many fundamental questions, all tied together, remain to be answered:
a) do we have continuous star formation or separate episodes, b) are all Milky Way GCs formed in the same way, c) and do they differ from those in dwarf galaxies, d) can we find very young GC-analogs to study the early stages directly, e) is one class of polluters enough to explain all GC properties,
f) what are the best diagnostic on which to focus?

\begin{acknowledgements}
AB thanks the organizers of the Workshop for their invitation.
We all thank Franca D'Antona for her seminal work and strong influence 
on the topic of multiple populations in clusters.
We acknowledge funding from the PRIN INAF 2009 ``Formation and Early Evolution of Massive Star Clusters".
\end{acknowledgements}

\bibliographystyle{aa}

\end{document}